\documentclass[12pt]{article}
\begin{document}
\def\ba{\begin{eqnarray}}
\def\ea{\end{eqnarray}}
\def\nn{\nonumber \\}

\title{Hydrodynamics and long range correlations}

\author{A.Bialas and K.Zalewski \\ H.Niewodniczanski Institute of
Nuclear Physics\\ Polish Academy of Sciences\thanks{Address: Radzikowskiego
152, 31-342 Krakow, Poland}\\and\\ M.Smoluchowski Institute of Physics
\\Jagellonian University\thanks{Address: Reymonta 4, 30-059 Krakow, Poland;
e-mail:bialas@th.if.uj.edu.pl;}}
 \maketitle

PACS: 25.75.Dw, 25.75.Gz

keywords: long-range correlations, hydrodynamic expansion

\begin{abstract}

It is shown that the recently proposed method of studying the long-range
correlations in multiparticle production can be effectively used to verify the
hydrodynamic nature of the longitudinal expansion  of the partonic system
created in the collision. The case of ALICE detector is explicitly considered.

\end{abstract}

\section{Introduction}

It is now commonly accepted that  particle production in high energy heavy ion
collisions  can be understood as a three step process: (i) the collision
creates a small and very dense state of matter which later (ii) undergoes the
hydrodynamic expansion and finally (iii) decays into observed hadrons. This
picture is well tested in the central rapidity region, where  the  hydrodynamic
evolution was studied in great detail \cite{florkowski}  and  where most of the
data were accumulated. The hydrodynamic description implies also, however,
longitudinal expansion of the system, as was  recognized already long  ago
\cite{florkowski}. Such picture is realized, e.g. in the well-known Landau
\cite{LAN} and Bjorken \cite{bj} models\footnote{See also \cite{abp}.}. Also
models based on the idea of saturation lead to a similar picture
\cite{CGC,CGC1}. It seems therefore of interest to investigate  consequences of
this observation. In the present note we point out that studies of  long-range
correlations in rapidity provide an effective tool for (a) testing the validity
of the hydrodynamic picture, as applied to longitudinal expansion and (b) to
uncover gross features of the  structure of the created system just before
freeze-out.

Our idea stems from the observation that  the system created in a high-energy
collision, after hydrodynamic evolution, decays into observed particles
locally, and thus  without introducing additional correlations, except the
short-range ones (as seen, e.g. in studies of the balance functions \cite{STAR,
BAL}). The observed long-range correlations in rapidity \cite{KIW,ABE,PHO} are
thus basically created by fluctuations in the total multiplicity. We argue
below that the detailed studies of these long-range correlations allow to
verify  this property.
 This feature is, admittedly, not specific to the hydrodynamic picture but since it is by far the
most developed  one and represents the most likely mechanism of the soft
particle production in heavy ion collisions,   we shall  use it in further
discussion. At the same time it should be kept in mind that, as discussed in
detail in our previous work \cite{BIZ1,BIZ2,BIZ3}, there are many other models
where this property is not satisfied\footnote{One example is the wounded
nucleon model  \cite{BIABC} which was shown \cite{BZW} to describe correctly
the data from \cite{PHO}.}. It is therefore necessary to clarify the situation.

Recently, we have developed \cite{BIZ1,BIZ2,BIZ3} a method of studying the
multiparticle long range correlations  occuring in high-energy particle
production processes . The method requires to measure  moments  of multiplicity
distribution in several bins, well separated in rapidity, as well as the joint
moments involving these bins\footnote{The method generalizes some previous work
in this direction, see \cite{ABE,BZD,LAL}.}. It was shown that such
investigations allow to test predictions of various models of particle
production in a rather general way, independent of a particular parametrization
of the model in question. In the present paper we show how the method can be
applied to studies of the longitudinal hydrodynamic expansion of the partonic
system produced in high-energy collisions.

The large rapidity interval available at the LHC makes the studies of long
range correlations a very attractive possibility. Unfortunately, all three
general purpose detectors are concentrated in the central rapidity region
(where the particle multiplicities can be measured with  good precision), while
the studies  of long range correlations demand also  detectors located off this
region.
 Some of them are under construction but at the moment (as far as we know) only the ALICE detector has two multiplicity counters located symmetrically $\pm 4$ units of rapidity from the center.
We shall argue that this situation gives a real possibility to verify the idea
of the hydrodynamic expansion of the parton system created at the early stages
of the collision.

 In  the ALICE experiment  measurements can be performed in three bins.
If one restricts oneself to  the  moments of order 3 (the errors for  higher
 moments are probably too large), and to symmetric collisions (like $pp$ or $PbPb$) one is left with 12 independent moments to be measured (see e.g. Eq. (\ref{ff}) . We shall show that when the  hydrodynamic description is accepted,  these 12 measured moments can be expressed in terms of 4 independent parameters, thus leading to 8 constraints to be verified experimentally.

There is, however, a practical problem which may appear when  such measurements
are undertaken. The space between the collision point and multiplicity
detectors located off the central rapidity region (which are of course crucial
for the success of the idea)   is filled  with a relatively large amount of
material \cite{ar}. Thus the particles produced at the collision point may
interact in this material before entering the counters. Consequently one may
expect that the particles created in these secondary interactions may form a
rather important background and thus bias  seriously the results.

In the present paper we discuss this problem and show that, even in the
presence of a substantial background of the type discussed above, the method
can still be useful in testing the hydrodynamic picture. In fact we shall show
that taking into account such background introduces only 3 more parameters in
the description of the measured moments.  This surely  makes the analysis more
complicated and reduces the number of constraints to 5, but  does not prevent
to obtain  conclusions.

In the next section we summarize briefly the method developed in \cite{BIZ2},
as applied to the present problem. In Section 3  the method  of background
treatment is explained.  Our conclusions are summarized in the last section.

\section {Multibin correlations}

In this section we show how the general method proposed in  \cite{BIZ2,BIZ3}
can be applied to the problem which interests us here.

Consider  multiplicity measurements in three bins "left", right" and "central",
denoted by $L,R,$ and $C$, respectively. To study long range correlations we
have to evaluate the joint probability to find the emitted particles in these
three bins.

In the hydrodynamic picture the observed hadrons are emitted from one "source"
which evolved  from the initial state created just after collision. The
long-range rapidity correlations in such system can only appear if present
already at the early stages of the collisions, before hydrodynamics comes into
play\footnote{Resonance decays can create some short-range correlations.}. They
are reflected in the total  multiplicity distribution $P(n)$ of the emitted
particles which can be usefully summarized by the generating function \ba
\Phi(z)=\sum_n P(n)z^n   \label{gen} \ea The next question to be asked is how
these particles are distributed among the three bins where the measurements are
performed. Assuming that this distribution  is completely random, i.e. it does
not introduce (nor erase)   any long-range correlations\footnote{This natural
assumption is generally used in studies of long-range rapidity correlations,
see e.g. \cite{BZD,CHY}.}, one obtains (for each $n$)  the multinomial
distribution \ba P(n_L,n_C,n_R;n)= \frac{n!}{n_L!n_C!n_R!}
p_L^{n_L}p_C^{n_C}p_R^{n_R}  \label{bern} \ea where $p_L,p_R$ and $p_C$ are the
probabilities for a particle to fall into the corresponding bin\footnote{Only
particles which populate the bins in question are considered, thus
$n_L+n_C+n_R=n$.}. These probabilities depend, of course, on the  rapidity
distribution of the emitted particles. They satisfy the obvious condition \ba
p_L+p_R+p_C=1 \ea For symmetric processes we have, naturally, $p_L=p_R\equiv p$
and thus $p_C=1-2p$. Showing that in this case there is just one probability to
be determined.

Using (\ref{gen}) and (\ref{bern}) one derives the generating function for the
particle distribution in the three bins \ba \phi(z_L,z_C,z_R)=
\Phi(p_Lz_L+p_Cz_C+p_Rz_R)  \label{gen3} \ea from which it is easy to derive
the expressions for the  12 measurable moments: \ba
F_{100}=F_{001}=pF_1\;;\;\;\; F_{010}=p_CF_1\;;\;\;\;
F_{200}=F_{002}=p^2F_2\;;\;\;\; F_{020}=p_C^2F_2;  \nn
F_{110}=F_{011}=pp_CF_2\;;\;\;\;F_{101}=p^2F_2\;;\;\;\;
F_{210}=F_{012}=p^2p_CF_3\;;\nn F_{120}=F_{021}=pp_C^2F_3\;;\;\;\;
F_{102}=F_{201}=p^3F_3\;;\;\;\;F_{111}=p^2p_CF_3\nn
F_{300}=F_{003}=p^3F_3\;;\;\;\; F_{030}=p_C^3F_3;  \label{ff} \ea where
\begin{equation}\label{}
F_{i_L,i_C,i_R} = \left\langle\frac{n_L!}{(n_L-i_L)!}\frac{n_C!}{(n_C-i_C)!}\frac{n_R!}{(n_R-i_R)!}\right\rangle
\end{equation}
and $n_b$ is the number of particles in bin $b$. One sees that all 12 moments
can  be expressed in terms of 4 parameters: the probability $p$ and the first
three moments of the total multiplicity distribution \ba F_i=\frac{d^i
\Phi(z)}{dz^i} [z=1]=\left<\frac{n!}{(n-i)!}\right>  \label{f} \ea One also
sees that already the two moments of rank 1 are sufficient to determine $p$ and
$F_1$. Furthermore, from   $F_{020}$ and $F_{030}$ one can determine $F_2$ and
$F_3$. Thus all other 8 moments can be predicted. Particularly useful seem to
be $F_{110}, F_{101}, F_{120}$ and $ F_{111}$  because they  involve only the
first powers of multiplicities in the left and right bins, where the
measurements are more difficult and  thus least accurate. This applies to an
idealized experiment. As mentioned in the Introduction, in practice one should
include some background corrections. This can be done, as explained in the
following section, but the number of predictions gets reduced. In particular,
to obtain any predictions at all, one should have more than two bins.

We conclude that  a good measurement of multiplicity distribution in the
central detector and a possibility to measure various (weighted) average
multiplicities in the left and right counters allow to test efficiently the
validity of mechanism of hydrodynamical expansion.

\section{Background in the left and right detectors}

As explained in the introduction, the signals observed  in the left and in the
right counters can be contaminated by
 products of secondary interactions of  particles produced at the interaction point when they pass through the material before arriving to counters. In the present section we show that this problem introduces some complications but does not prevent effective tests of the hydrodynamic evolution.

The basic observation is that the background due to the secondary interactions
modifies the generating function (\ref{gen3})  in a straightforward  way: one
simply  has to perform the replacement \ba z_L\rightarrow
h(z_L)\;;\;\;\;\;z_R\rightarrow h(z_R)  \label{repl} \ea where \ba h(z)=\sum_m
w(m) z^m \ea is the generating function describing the probability distribution
$w(m)$ of particles induced  by one of the particles produced in the collision
and passing to one (left or right) of the counters. With this substitution the
generating function of the observed distribution is \ba
\Phi(z_L,z_C,z_R)=\Phi[p\;h(z_L) + p_Cz_C + p\;h(z_R)] . \label{fib} \ea Since
the moments of rank $r$ are expressed by derivatives of (\ref{fib}) up to order
$r$, one sees that the replacement (\ref{repl}) introduces three new
parameters. This number may perhaps be  reduced if more information on the
properties of the background contribution can be extracted from data. However,
even with this most general formulation significant tests can be performed, as
we discuss below.

Indeed, it is clear that, if one restricts to moments up to the third order,
only three moments of the distribution (\ref{}) are needed. Thus altogether
one needs 7 parameters to evaluate 12 moments of the distribution which can be
measured and  one obtains 5 independent constraints which can be used to the
test the model. This can be explicitly  seen from
 the formulae for the moments derived from  the generating function (\ref{fib})
\ba F_{100}=F_{001}=p_1F_1\;;\;\;\; F_{010}=p_CF_1\;;\;\;\;
F_{200}=F_{002}=p_2F_1+p_1^2F_2\;;\nn F_{020}=p_C^2F_2\;;\;\;\;
F_{110}=F_{011}=p_1p_CF_2\;;\;\;\;F_{101}=p_1^2F_2;\nn
F_{210}=F_{012}=p_2p_CF_2+p_1^2p_CF_3\;;\;\;\;F_{120}=F_{021}=p_1p_C^2F_3\nn
F_{102}=F_{201}=p_1p_2F_2+p_1^3F_3\;;\;\;\;F_{111}=p_1^2p_CF_3\nn
F_{300}=F_{003}=p_3F_1+3p_2p_1F_2+ p_1^3F_3\;;\;\;\; F_{030}=p_C^3F_3;
\label{ffb} \ea where $F_i$ are given by (\ref{f}) and $p_i=p\;h_i$ with \ba
h_i\equiv \frac{d^ih(z)}{dz^i}[z=1] =\left<\frac {m!}{(m-i)!}\right>  \label{h}
\ea being the factorial moments of the distribution $w(m)$.

Although the formulae (\ref{ffb}) are more complicated than (\ref{ff}), they
are still tractable. Indeed, one sees that from the measured
$F_{100},F_{010},F_{020}$ and $F_{030}$  one can determine  $F_{110}$,
$F_{101},F_{120}$ and $F_{111}$, i.e. all moments which require only
measurement of first powers of multiplicities in the left and right counters.
If the second order moments can also be measured in these counters, one more
relation can be tested.  Measurement of $F_{300}$ yields the parameter $p_3$
but does not provide any new relation.

\section{Summary and comments}

The recently developed method of studying the multiparticle long-range
correlations was applied to correlations in rapidity which arise when the
parton system created in a high-energy collision undergoes  hydrodynamic
expansion in the  longitudinal direction.

It was  shown that the measurements of multiplicity distributions in three
symmetrically positioned rapidity  bins,  separated by a sufficient distance
(say, 3-4 units of rapidity) give a practical possibility to test the validity
of the idea that the hydrodynamic expansion dominates the longitudinal
evolution of the parton system created in a high-energy collision. It was also
argued that such measurements seem feasible in the ALICE detector, even in
presence of important background in the forward and backward multiplicity
counters.

The following comment are in order.

(i) The idea to study the long-range correlations in more than two bins was
pioneered in the recent paper by STAR collaboration \cite{ABE}. The
measurements reported in this paper show that the dual parton model \cite{CST},
being closest to data,  predicts nevertheless the correlation strength
somewhat below the observations. As discussed in \cite{BIZ1,BIZ2} such a model
(which comprises 3 kinds of particle sources) gives weaker long-range
correlations than the hydro picture discussed here (where there are only
sources of one kind). One may therefore hope that, as expected, the hydro
evolution will provide a better description. On the other hand, the data from
PHOBOS \cite{PHO} were shown  \cite{BZW} to be consistent  with the wounded
nucleon model \cite{BIA}. Thus the situation is somewhat confusing and a more
systematic approach, as proposed here, seems indeed necessary for its
clarification.

(ii) Since the UA5 data on particle production in $pp$ collisions \cite{UA5}
indicate significant deviations from the hydrodynamic picture  \cite{BZD},
comparison of $pp$ and $PbPb$ collisions may be of particular interest, as some
significant differences may be expected if the longitudinal hydrodynamic
expansion does indeed dominate in heavy ion collisions.

(iii) The method is not restricted to the minimum bias sample. It can be used
to study e.g. what happens when a high-multiplicity trigger is applied in $pp$
collisions. This could allow to study to what extent the high multiplicity $pp$
collisions are close to heavy ion collisions (as suggested by the recent
discovery of the "ridge" in high multiplicity pp collisions by the CMS coll.
\cite{CMS}).

(iv)  As seen from (\ref{ffb}), to obtain a meaningful test under  the
conditions of the ALICE experiment, it is necessary to measure moments at least
up to rank 3. If, as expected,  the test is fulfilled  for $PbPb$ data, one
obtains in addition important  information on the parameters which describe
the system close to freeze-out. From that point of view it would be desirable
to measure the moments of even higher rank.

(v)  If the test fails, this would indicate a nontrivial internal structure in
the system (e.g. several independent particle sources), not considered till now
in the hydro calculations (although suggested in some old models
\cite{BIABC,CST,BIA}).

To conclude, we feel that since the idea of hydrodynamic evolution presently
dominates  the  description of particle production at high energies,
particularly for heavy ion collisions,
 the investigation of long range correlations in rapidity we propose may help to understand the dynamics of these processes.

\section{Acknowledgements}
We would like to thank Andrzej Rybicki for conversations which led us to
undertake this investigation. This work was supported in part by the grant N
N202 125437 of the Polish Ministry od Science and Higher Education (2009-2012).

\end{document}